\documentstyle[epsfig]{aipproc}

\begin{document}

%%%%%%%%%%%%%%%%%%%%% Title %%%%%%%%%%%%%%%%%%%%%%

\title{Nucleon resonances in polarized $\bbox{\omega}$ photoproduction}

%%%%%%%%%%%%%%%%%%%% Authors %%%%%%%%%%%%%%%%%%%%%

\author{
Yongseok Oh\,$^a$, Alexander I. Titov\,$^{b}$, and T.-S. H. Lee\,$^c$}

%%%%%%%%%%%%%%%%%%%% Addresses %%%%%%%%%%%%%%%%%%%%%

\address{
$^a$
Institute of Physics and Applied Physics,
Yonsei University, Seoul 120-749, Korea\\
$^b$
Bogolyubov Laboratory of Theoretical Physics, JINR,
Dubna 141980, Russia\\
$^c$
Physics Division, Argonne National Laboratory, Argonne,
Illinois 60439}

\maketitle

%%%%%%%%%%%%%%%%%%%% Abstract %%%%%%%%%%%%%%%%%%%%%

\begin{abstract}

The role of the nucleon resonances ($N^*$) in $\omega$ photoproduction
is investigated by using the resonance parameters predicted by Capstick
and Roberts.
The contributions from the nucleon resonances are found to be significant
in various spin asymmetries.
In particular, we found that a crucial test of our predictions can be
made by measuring the parity asymmetry and beam-target double asymmetry 
at forward scattering angles. 

\end{abstract}

The constituent quark models predict a much richer nucleon excitation
spectrum than what has been observed in pion-nucleon scattering
\cite{IK77-80}.
This has been attributed to the possibility that a lot of the predicted
nucleon resonances ($N^*$) could couple weakly to the $\pi N$ channel
\cite{Caps00}.
Therefore it is necessary to search for the nucleon excitations in other
reactions to resolve the so-called ``missing resonance problem.''
Electromagnetic production of vector mesons ($\omega,\rho,\phi$) is one
of such reactions and is being investigated experimentally, e.g.,
at LEPS of SPring-8, TJNAF, ELSA-SAPHIR of Bonn and GRAAL of Grenoble.

The role of the nucleon excitations in vector meson photoproduction was
studied recently by Zhao {\em et al.\/} \cite{ZLB98,ZDGS99} within 
an $\mbox{SU}(6) \times \mbox{O}(3)$ constituent quark model.
With the meson-quark coupling parameters adjusted to fit the existing
data, they found that the single polarization observables are sensitive
to the nucleon resonances.

We are motivated by the predictions by Capstick and Roberts
\cite{Caps92,CR94} based on the constituent quark model which accounts
for the configuration mixing due to the residual quark-quark interactions
\cite{GI85-CI86} and the ${}^3P_0$ model~\cite{Yaou88} for the meson
decay channels.
Thus it would be interesting to see how these predictions differ from
those of Refs. \cite{ZLB98,ZDGS99} and whether it can be tested
experimentally.

We focus on $\omega$ photoproduction in this work, simply because
its non-resonant reaction mechanisms are fairly well established
\cite{JLMS77,FS96}.
This reaction is dominated by diffractive process at high energies
and by one-pion exchange at low energies which may be assumed as the
dominant part of non-resonant background and may be used as a starting
point for investigating the $N^*$ effects.

We assume that the non-resonant (background) invariant amplitude has
the form
\begin{eqnarray}
I^{bg}_{fi} = I^P_{fi} + I^{ps}_{fi} + I^{N}_{fi},
\end{eqnarray}
where $I^P_{fi}$, $I^{ps}_{fi}$, and $I^{N}_{fi}$ denote the
amplitudes due to the Pomeron, pseudoscalar-meson exchange, and
direct and crossed nucleon terms, respectively.
The four-momenta of the incoming photon, outgoing $\omega$, initial
nucleon, and final nucleon are denoted as $k$, $q$, $p$, and $p'$
respectively, which defines $t = (p - p')^2 = (q-k)^2$,
$s \equiv W^2 = (p+k)^2$, and the $\omega$ production angle $\theta$
by $\cos\theta \equiv {\bf k} \cdot {\bf q} / |{\bf k}| |{\bf q}|$.

For the Pomeron exchange, which governs the total cross sections and
differential cross sections at low $|t|$ in the high energy region,
we follow the Donnachie-Landshoff model \cite{DL84-92}.
For the details of this model, we refer to, e.g., Refs.~\cite{LM95,PL97}.
The pseudoscalar-meson exchange amplitude is calculated from the effective
Lagrangian of Refs.~\cite{JLMS77,FS96} with slightly modified
cut-off parameters $\Lambda_\pi = 0.6$ GeV and
$\Lambda_{\omega\gamma\pi} = 0.7$ GeV.

We evaluate the direct and crossed nucleon terms from the 
Lagrangian,
\begin{equation}
{\cal L}_{V PP}^{}  = 
- g_V^{} \bar{P} \left( \gamma_\mu  {V}^\mu -\frac{\kappa_V^{}}{2m_p^{}}
\sigma^{\mu\nu} \partial_\nu V_\mu \right) P,
\end{equation}
where $P$ stands for the proton Dirac spinor and $V$ denotes $\gamma$ or
$\omega$.
When $V=\gamma$, we have $g_{\gamma}=e$ and $\kappa_{\gamma} = 1.79$.
%which is the anomalous magnetic moment of the nucleon.
For the $\omega NN$ coupling, i.e., when $V = \omega$, we take
$g_{\omega NN}^{} = 10.35$ and $\kappa_\omega=0$, which are determined in
a study of $\pi N$ scattering and pion photoproduction \cite{SL96}.
The $\omega NN$ vertices are dressed by the form factor,
$\Lambda_N^4/[\Lambda_N^4 + (p^2-m_p^2)^2]$, where $p$ is
the four momentum of the off-shell nucleon with $\Lambda_N=0.5$ GeV.

In order to estimate the nucleon resonance contributions we make use
of the quark model predictions on the resonance photo-excitation
($\gamma N \to N^*$) and the resonance decay ($N^* \to \omega N$)
reported in Refs. \cite{Caps92,CR94} using a relativised quark model. 
Referring the detailed description of our resonant model to Ref.
\cite{OTL00}, here we discuss the main results of our investigation.
The resonant amplitude is defined via $N^*$ production amplitude
${\cal M}_{\gamma N \to N^*}$ and decay amplitude ${\cal M}_{N^*\to
N'\omega}$:
\begin{equation}
I^{N^*}
\propto \sum_{J,M_J^{}}
{\cal M}_{N^*\to N'\omega}
{\cal M}_{\gamma N \to N^*}
\,/\,\left({\sqrt{s} - M_R^J + \frac{i}{2}\Gamma^J(s)}\right),
\label{T:N*}
\end{equation}
where $M^J_R$ is the mass of an $N^*$ with spin quantum numbers
$(J, M_J)$ and $\Gamma^J(s)$ is the energy dependent total decay width
\cite{YSAL00}.
Since the most nucleon resonances we are dealing with are missing
resonances, there is no information for their total decay widths.
Therefore we rely on the averaged decay widths of $N^*$ listed in
Particle Data Group \cite{PDG00} and take $\Gamma^J(M_R^J) \simeq 300$
MeV.
The amplitudes ${\cal M}_{\gamma N \to N^*}$ and
${\cal M}_{N^*\to N'\omega}$ are related to the corresponding transition
amplitude as ${\cal M}_{\gamma N \to N^*} \propto  A_{M_J^{}}^{}$ and
${\cal M}_{N^*\to N'\omega} \propto \sum G(J,L,S)$, where the resonance
parameters are taken from Refs. \cite{Caps92,CR94}.
In this study, we consider 12 positive parity and 10 negative parity
nucleon resonances up to spin-$9/2$.
Three of them were seen in the $\pi N$ channel with four-star rating,
five of them with two-star rating, and one of them with one-star rating.
(See Ref. \cite{OTL00}.)
The majority of the predicted $N^*$'s are ``missing'' so far.
Here we should also mention that we are not able to account for the
resonances with the predicted masses less than the $\omega N$ threshold,
since their decay vertex functions with an off-shell momentum are
not available yet in the model of Refs. \cite{Caps92,CR94}.

\begin{figure}[t]
\centering
\epsfig{file= 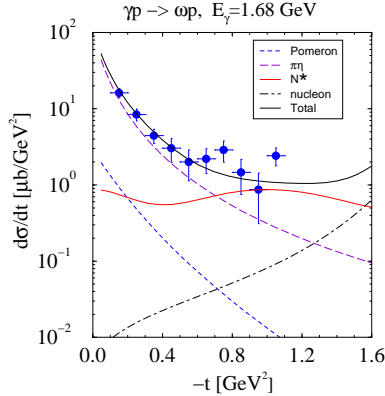, width=5cm}
\caption{
Differential cross sections of $\gamma p\to p\omega$ reaction
as a function of $t$ at $E_\gamma = 1.68$ GeV.
Data are taken from Ref. \protect\cite{Klein96-98}.}
\label{fig:dsdt}
\end{figure}

As an example for the role of nucleon resonances, we present our
results for the differential cross section of $\omega$ photoproduction
at $E_\gamma=1.68$ GeV in Fig.~\ref{fig:dsdt}.
We also found that the data could be described to a very large extent
for $E_\gamma \leq 5$ GeV.
One can see that the contributions due to the $N^*$ excitations (dotted
line) and the direct and crossed nucleon terms (dot-dashed line)
help bring the agreement with the data at large angles.
Close inspection of the resonance part shows that the contributions from
$N\frac32^+ (1910)$ and $N\frac32^- (1960)$ are the largest at 
$W=1.79 \sim 2.12$ GeV.
In Ref.~\cite{Caps92}, the $N\frac32^- (1960)$ is identified as a two
star $D_{13}(2080)$ resonance of PDG, while the $N\frac32^+ (1910)$ is a
missing resonance.
In the study of Ref. \cite{ZLB98}, the authors found that
$F_{15}(2000)$ dominates.
This resonance is identified with $N{\frac52^+}(1995)$ in Ref.
\cite{Caps92} and is found to be not so strong in our calculation.
The difference between the two calculations reflects the difference
of the employed quark models.

\begin{figure}[t]
\centering
\epsfig{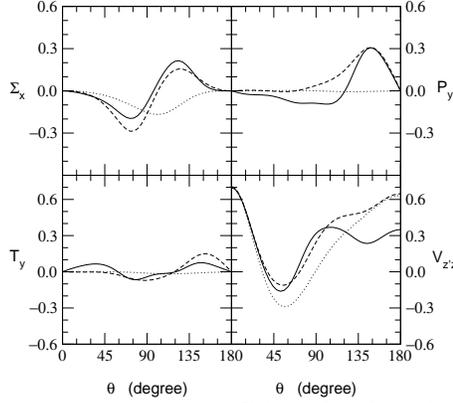}
\caption{
Single asymmetries at $E_\gamma = 1.7$ GeV.
The dotted curves are calculated without including $N^*$ effects, the
dashed curves include contributions of $N\frac32^+(1910)$ and
$N\frac32^-(1960)$ only, and the solid curves are calculated with all
$N^*$.}
\label{fig:single}
\end{figure}

Since it is difficult to test our predictions by considering only the
angular distributions, we turn to the spin variables.
We first examine the single spin observables~\cite{TOYM98,PST96}.
Our predictions for photon asymmetry ($\Sigma_x$), target asymmetry
($T_y$), recoil nucleon asymmetry ($P_y$), and vector-meson tensor
asymmetry ($V_{z'z'}$) are shown in Fig.~\ref{fig:single}.
%The $T_y$ and $P_y$ asymmetries are proportional to $\sin\delta_{kl}$,
%where $\delta_{kl}$ is the relative phase of the different amplitudes.
%For the background part this sum deviates from zero through the
%interference between the Pomeron exchange
%and other parts of background amplitude which is extremely small.
%That is,  for the non-resonant background we expect 
%$T_y$, $P_y$=0.
%For pure pseudoscalar exchange $\Sigma_x=0$. Moreover
%it is equal zero for $\theta=0,\pi$ for all channels.
We find that the $N^*$ excitations change the predictions from the
dotted curves to the solid curves.
The dashed curves are obtained when only the $N\frac32^+(1910)$ and
$N\frac32^-(1960)$ are included in calculating the resonant part of the
amplitude. 
Although our predictions are different from those of Ref. \cite{ZLB98},
we confirm their conclusion that the single polarization
observables are sensitive to the $N^*$ excitations but mostly at large
scattering angles.

\begin{figure}[t]
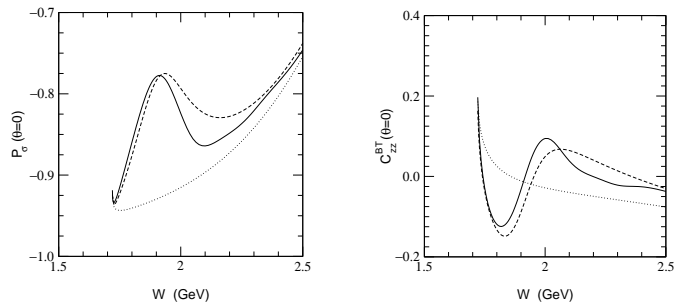

\centering
\epsfig{file=OHfig3l.eps, width=4cm}\qquad
\epsfig{file=OHfig3r.eps, width=4cm}
\caption{
Parity asymmetry $P_\sigma$ at $\theta=0$ (left panel) and beam-target
asymmetry $C^{\rm BT}$ (right panel) as a function of $W$.
Notations are the same as in Fig.~\ref{fig:single}.}
\label{fig:Psigma}
\end{figure}

In order to probe the role of the nucleon resonances in $\omega$
photoproduction, we address two polarization observables that are
sensitive to the $N^*$ contributions {\em at forward scattering angles\/}.
The first one is the parity asymmetry $P_\sigma$~\cite{SSW}.
At forward scattering region where the one-pion exchange is dominant,
one expects $P_\sigma = -1$. 
Thus any deviation from this value will be only due to $N^*$ excitation
and Pomeron exchange, since the contribution from the direct and crossed
nucleon terms is two or three orders in magnitude smaller at
$\theta = 0$ (see Fig.~\ref{fig:dsdt}).
Our predictions for $P_\sigma$ are shown in Fig.~\ref{fig:Psigma}
(left panel). We show the results from calculations with (solid curve) 
and without (dotted curve) including the $N^*$ contributions.
The difference between them is striking and can be unambiguously tested
experimentally.
Here we also find that the $N\frac32^+ (1910)$ and $N\frac32^- (1960)$
contributions are dominant.
By keeping only these two resonances in calculating the resonant part of
the amplitude, we obtain the dashed curve which is not too different from
the full calculation (solid curve).
Another asymmetry which is sensitive to the $N^*$ excitations at forward
scattering angles is the beam-target double asymmetry ($C^{BT}_{zz}$)
\cite{TOYM98}.
Given in the right panel of Fig.~\ref{fig:Psigma} are our predictions on
$C^{BT}_{zz}$ at $\theta=0$ as a function of invariant mass $W$.
The striking difference between the solid curve and dotted curve is due to
the $N^*$ excitations.
Again, the $N\frac32^+ (1910)$ and $N\frac32^- (1960)$ give the dominant
contributions (dashed curve).

In summary, we have investigated the role of nucleon resonances in
$\omega$ photoproduction, especially in the resonance region.
It was found that their role is important in the differential cross
sections at large angles and some spin asymmetries can be used to
identify the role of the nucleon resonances at forward scattering angles
where precise measurements might be more favorable because the cross
sections are peaked at $\theta = 0$.
Experimental test of them will be a useful step toward resolving the
so-called ``missing resonance problem'' or distinguishing
different quark model predictions.

\bigskip
\noindent
{\bf Acknowledgements}.
This work was supported in part by the Brain Korea 21 project of Korean
Ministry of Education, Russian Foundation for Basic Research under 
Grant No. 96-15-96426, and U.S. DOE Nuclear Physics Division Contract
No. W-31-109-ENG-38.

%%%%%%%%%%%%%%%%%  References  %%%%%%%%%%%%%%%%%%%%%%%

\end{document}